\documentclass[conference]{IEEEtran}
\IEEEoverridecommandlockouts
\usepackage{cite}
\usepackage{amsmath,amssymb,amsfonts}
\usepackage{amssymb}
\usepackage{algpseudocode}
\usepackage{amsfonts}
\usepackage{graphicx}
\usepackage{fancyhdr}  %
\usepackage{cases}
\usepackage{textcomp}
\usepackage{extarrows}
\usepackage{algorithm,algpseudocode}
\usepackage{multirow,tabularx}
\usepackage{mathtools}
\usepackage{xcolor}
\usepackage[english]{babel}
\usepackage{amsthm}

\newtheorem*{remark}{Remark}

\def\BibTeX{{\rm B\kern-.05em{\sc i\kern-.025em b}\kern-.08em
    T\kern-.1667em\lower.7ex\hbox{E}\kern-.125emX}}
\begin{document}

%\title{Iterative Channel Estimation for RIS-aided Wireless Communication Systems
\title{Parallel Factor Decomposition Channel Estimation in RIS-Assisted Multi-User MISO Communication
%\title{Iterative Channel Estimation for RIS-Assisted Multi-User MISO Communication using the PARAFAC Decomposition
%\thanks{Identify applicable funding agency here. If none, delete this.}
}

\author{
\IEEEauthorblockN{Li Wei$^1$, Chongwen Huang$^1$, George~C.~Alexandropoulos$^2$, and Chau Yuen$^1$
 }
\IEEEauthorblockA{
$^1$Engineering Product Development, Singapore University of Technology and Design, Singapore\\
$^2$Department of Informatics and Telecommunications, National and Kapodistrian University of Athens, Greece\\
emails: wei\_li@mymail.sutd.edu.sg, \{chongwen\_huang, yuenchau\}@sutd.edu.sg, alexandg@di.uoa.gr
}}

%\author{
%\begin{tabular}[t]{c@{\extracolsep{5em}}c}
%Li Wei  & Chongwen Huang \\
%\textit{Engineering Product Development} & \textit{Engineering Product Development} \\
%\textit{Singapore University of Technology and Design} & \textit{Singapore University of Technology and Design} \\
%wei\_li@mymail.sutd.edu.sg & chongwen\_huang@sutd.edu.sg
%\end{tabular} \\
%\begin{tabular}[t]{c@{\extracolsep{5em}}c}
%George~C.~Alexandropoulos  & Chau Yuen \\
%\textit{Department of Informatics and Telecommunications} & \textit{Engineering Product Development }\\
%\textit{National and Kapodistrian University of Athens} & \textit{Singapore University of Technology and Design }\\
%alexandg@di.uoa.gr & yuenchau@sutd.edu.sg
%\end{tabular}
%}
%\author{
%\hspace{-30mm}\IEEEauthorblockN{Li Wei}
%\IEEEauthorblockA{\hspace{-30mm}\textit{Engineering Product Development} \\
%\hspace{-30mm}\textit{Singapore University of Technology and Design}\\
%\hspace{-30mm}wei\_li@mymail.sutd.edu.sg}
%\and
%\IEEEauthorblockN{Chongwen Huang}
%\IEEEauthorblockA{\textit{Engineering Product Development} \\
%\textit{Singapore University of Technology and Design}\\
%chongwen\_huang@sutd.edu.sg}
%\and
%\hspace{6mm}\IEEEauthorblockN{George~C.~Alexandropoulos}
%\hspace{6mm}\IEEEauthorblockA{\hspace{6mm}\textit{Department of Informatics and Telecommunications} \\
%\hspace{6mm}\textit{National and Kapodistrian University of Athens}\\
%\hspace{6mm}alexandg@di.uoa.gr}
%\and
%\IEEEauthorblockN{Chau Yuen}
%\IEEEauthorblockA{\textit{Engineering Product Development} \\
%\textit{Singapore University of Technology and Design}\\
%yuenchau@sutd.edu.sg}
%}
\maketitle

\begin{abstract}
Reconfigurable Intelligent Surfaces (RISs) have been recently considered as an energy-efficient solution for future wireless networks due to their fast and low power configuration enabling massive connectivity and low latency communications. Channel estimation in RIS-based systems is one of the most critical challenges due to the large number of reflecting unit elements and their distinctive hardware constraints. In this paper, we focus on the downlink of a RIS-assisted multi-user Multiple Input Single Output (MISO) communication system and present a method based on the PARAllel FACtor (PARAFAC) decomposition to unfold the resulting cascaded channel model. The proposed method includes an alternating least squares algorithm to iteratively estimate the channel between the base station and RIS, as well as the channels between RIS and users. Our selective simulation results show that the proposed iterative channel estimation method outperforms a benchmark scheme using genie-aided information. We also provide insights on the impact of different RIS settings on the proposed algorithm.  
\end{abstract}


\section{Introduction}\label{sec:intro}
Reconfigurable Intelligent Surfaces (RISs) are lately considered as a candidate technology for beyond fifth Generation (5G) wireless communication due to their potential significant benefits in low powered, energy-efficient, high-speed, massive-connectivity, and low-latency communications\cite{Akyildiz2018mag,holobeamforming,hu2018beyond,huang2019reconfigurable,Marco2019,qingqing2019towards}. In its general form, a RIS is a reconfigurable planar metasurface composed of a large number of hardware-efficient passive reflecting elements \cite{Akyildiz2018mag,tang2019wireless,huang2018achievable}. Each unit element can alter the phase of the incoming signal without requiring a dedicated power amplifier, as for example needed in conventional amplify-and forward relaying systems \cite{Akyildiz2018mag,huang2018achievable,holobeamforming,huang2019reconfigurable,li2019irs}. 

The energy efficiency potential of RIS in the downlink of outdoor multi-user Multiple Input Single Output (MISO) communications was analyzed in \cite{huang2019reconfigurable}, while \cite{husha_LIS2} focused on an indoor scenario to illustrate the potential of RIS-based indoor positioning. It was shown in \cite{huang2018energy,han2019} that the latter potential is large even in the case of RISs with finite resolution unit elements and statistical channel knowledge. Recently, \cite{yan2019passive} proposed a novel passive beamforming and information transfer technique to enhance primary communication, and a two-step approach at the receiver to retrieve the information from both the transmitter and RIS. RIS-assisted communication in the millimeter wave \cite{wang2019intelligent} and THz \cite{ning2019intelligent} bands was also lately investigated. However, most of the existing research works focusing on beamforming and RIS phase matrix optimization techniques assumed perfect channel state information availability.  

Channel estimation in RIS-assisted multi-user communications is a challenging task due to the massive channel training overhead required \cite{liaskos2018new,liang2019large,huang2019holographic}. Particularly, the direct channels between the Base Station (BS) and each user, the channel between RIS and BS, and the channels between each user and RIS need to be estimated given that RIS is equipped with large numbers of unit elements having non-linear hardware characteristics. The recent works \cite{Alkhateeb2019} and \cite{huang2019spawc} presented compressive sensing and deep learning approaches for recovering the involved channels and designing the RIS phase matrix. However, the deep learning approaches require extensive training during offline training phases, and the compressive sensing framework assumes that RIS has some active elements, specifically, a fully digital or hybrid analog and digital architecture attached at a RIS portion. The latter architectures increase the RIS hardware complexity and power consumption. A low-power receive radio frequency chain for channel estimation was considered in \cite{qingqing2019towards}, which also requires additional energy consumption compared to passive RIS. Very recently, \cite{zheng2019intelligent} presented a technique for channel estimation in the frequency domain, however the separate RIS-based channels were not explicitly estimated. 

In this paper, motivated by the PARAllel FACtor (PARAFAC) decomposition \cite{rong2012channel, ximenes2014parafac}, we present an efficient channel estimation method for all involved channels in the downlink of a RIS-aided multi-user MISO system. PARAFAC is a method to simultaneously analyze matrix factors, which extends the standard two-way factor analysis to tensor decomposition \cite{harshman1994parafac,bro2003new,ten2002uniqueness,roemer2008closed}. Based on this decomposition, the proposed channel estimation method includes an Alternating Least Squares (ALS) algorithm to iteratively estimate at the receiver side the channel between BS and RIS, and the channels between RIS and each of the users by leveraging unfolded forms \cite{sidiropoulos2000blind}\cite{kolda2009tensor}. Representative simulation results validate the accuracy of the proposed estimation method and its superiority over conventional channel estimation. 
%The remainder of this paper is organized as follows. In Section \ref{sec:format}, the considered system model is given with the problem formulation. Details of PARAFAC-based channel estimation algorithm are provided in Section \ref{sec:channel_est}. Section~\ref{sec:simulation} presents the estimation performance under some considered scenarios. Finally, concluding remarks are drawn in Section~\ref{sec:conclusion}.

\textit{Notation}: Fonts $a$, $\mathbf{a}$, and $\mathbf{A}$ represent scalars, vectors, and matrices, respectively. $\mathbf{A}^T$, $\mathbf{A}^H$, $\mathbf{A}^{-1}$, $\mathbf{A^\dag}$, and $\|\mathbf{A}\|_F$ denote transpose, Hermitian (conjugate transpose), inverse, pseudo-inverse, and Frobenius norm of $ \mathbf{A} $, respectively.  $|\cdot|$ and $(\cdot)^*$ denote the modulus and conjugate, respectively. $\text{tr}(\cdot)$ gives the trace of a matrix and $\mathbf{I}_n$ (with $n\geq2$) is the $n\times n$ identity matrix. $\otimes$ and $\circ$ represent the Kronecker and the Khatri-Rao (column wise Kronecker) matrix products, respectively. Finally, notation $diag(\mathbf{a})$ represents a diagonal matrix with the entries of $\mathbf{a}$ on its main diagonal.

\section{System and Signal Models}\label{sec:format}
In this section, we first describe the system and signal models for the considered RIS-assisted multi-user MISO communication system, and then present the PARAFAC decomposition for the end-to-end wireless channel.
\subsection{System Model}\label{subsec:signal model}
\begin{figure}\vspace{-0mm}
	\begin{center}
		\centerline{\includegraphics[width=0.42\textwidth]{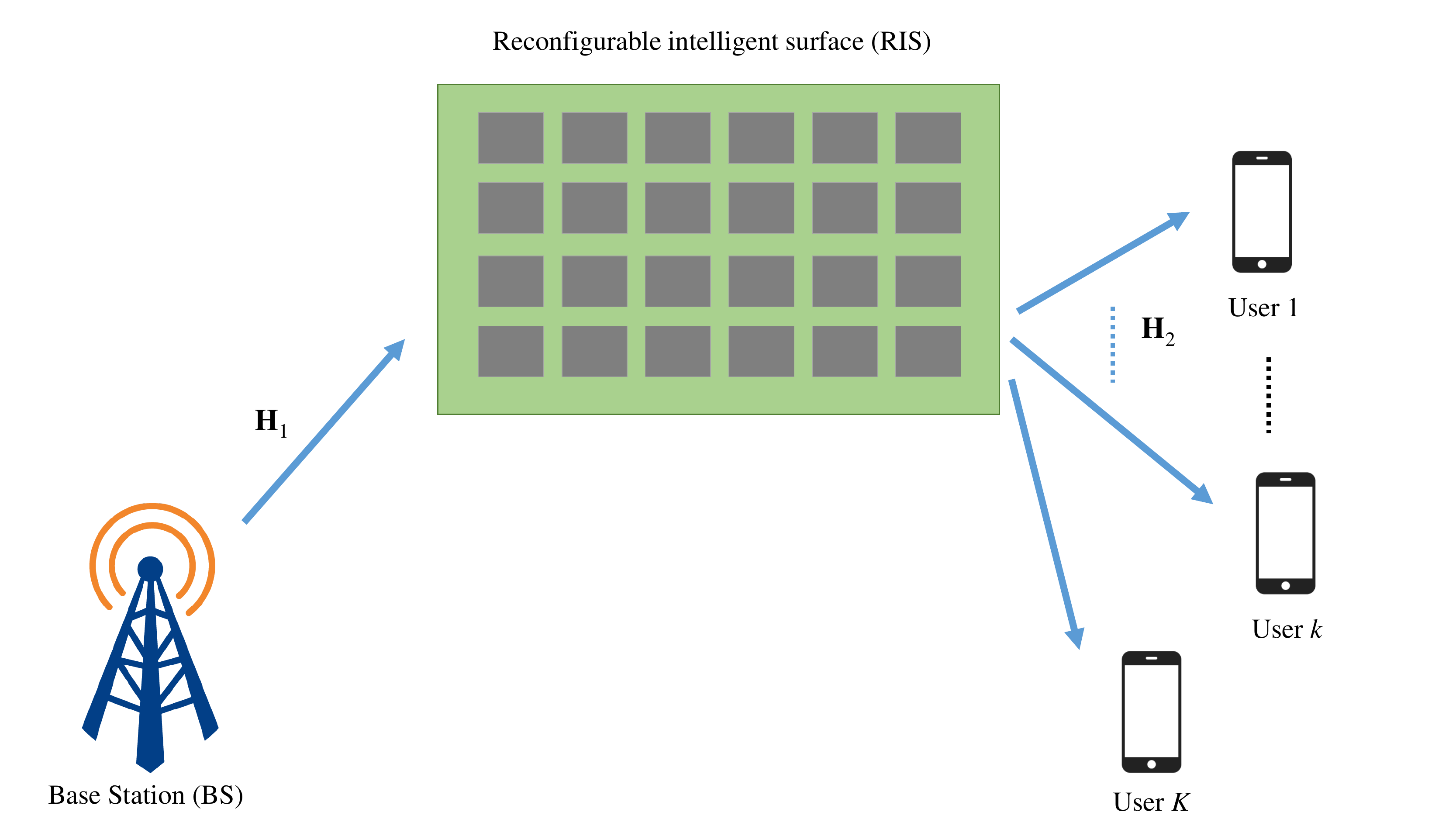}}  \vspace{-0mm}
		\caption{The considered RIS-based multi-user MISO system consisting of a $M$-antenna base station simultaneously serving in the downlink $K$ single-antenna mobile users.}
		\label{fig:Estimation_Scheme} \vspace{-4mm}
	\end{center}
\end{figure}

Consider the downlink communication between a BS equipped with $M$ antenna elements and $K$ single-antenna mobile users. We assume that this communication is realized via a discrete-element RIS \cite{holobeamforming} deployed on the facade of a building existing in the vicinity of the BS side, as illustrated in Fig$.$~\ref{fig:Estimation_Scheme}. The RIS is comprised of $N$ unit cells of equal small size, each made from metamaterials capable of adjusting their reflection coefficients. The direct signal path between the BS and the mobile users is neglected due to unfavorable propagation conditions, such as large obstacles. The received discrete-time signals at all $K$ mobile users for $T$ consecutive time slots using the $p$-th RIS phase configuration (out of the $P$ available in total, hence, $p=1,2,\ldots,P$) can be compactly expressed with $\mathbf{Y}_{p}\in\mathbb{C}^{K \times T}$ given by
\begin{equation}\label{equ:YXZ}
\mathbf{Y}_{p}\triangleq\mathbf{H}_{2} D_p(\mathbf{\Phi})  \mathbf{H}_{1} \mathbf{X}+\mathbf{W}_{p},
\end{equation}
where $D_p(\mathbf{\Phi})\triangleq diag (\mathbf{\Phi}_{p,:})$ with $\mathbf{\Phi}_{p,:}$ representing the $p$-th row of the $P\times N$ complex-valued matrix $\mathbf{\Phi}$. Each row of $\mathbf{\Phi}$ includes the phase configurations for the RIS unit elements, which are usually chosen for low resolution discrete sets \cite{huang2018energy}. $\mathbf{H}_{1}\in\mathbb{C}^{N\times M}$ and $\mathbf{H}_{2}\in\mathbb{C}^{K\times N}$ denote, respectively, the channel matrices between RIS and BS, and between all users and RIS. Additionally, $\mathbf{X}\in\mathbb{C}^{M \times T}$ includes the BS transmitted signal within $T$ time slots (it must hold $T \geq M$ for channel estimation), and $\mathbf {W}_p \in \mathbb{C}^{K \times T}$ is the Additive White Gaussian Noise (AWGN) matrix having zero mean and unit variance elements.

\subsection{Decomposition of the Received Training Signal}\label{subsec:prob_formu}
The channel matrices $\mathbf{H}_{1}$ and $\mathbf{H}_{2}$ in \eqref{fig:Estimation_Scheme} are in general unknown and need to be estimated. We hereinafter assume that these matrices have independent and identically distributed complex Gaussian entries; the entries between the matrices are also assumed independent. We further assume orthogonal pilot signals in \eqref{fig:Estimation_Scheme}, such that $\mathbf{X} \mathbf{X}^{H}=\mathbf{I}_{M}$, and that $\mathbf{\Phi}$ with the $P$ distinct configurations is used during the channel estimation phase. In order to estimate $\mathbf{H}_{1}$ and $\mathbf{H}_{2}$, we first apply the PARAFAC decomposition in \eqref{equ:YXZ}, according to which the intended channels are represented using tensors \cite{kroonenberg1980principal,rong2012channel}. In doing so, we define for each $p$-th RIS training configuration the matrix $\widetilde{\mathbf{Z}}_{p}\in\mathbb{C}^{K\times M}$ after removing the pilot symbols as
\begin{equation}\label{widetilde_Z_p}
\widetilde{\mathbf{Z}}_{p}\triangleq\mathbf{Y}_{p}\mathbf{X}^{H}=\mathbf{Z}_{p}+\widetilde{\mathbf{W}}_{p},
\end{equation}
where $\mathbf{Z}_{p} \in \mathbb{C}^{K \times M}$ is the noiseless version of the end-to-end RIS-based wireless channel given by
\begin{equation}\label{Z_p}
\begin{aligned}
\mathbf{Z}_{p}\triangleq\mathbf{H}_{2} D_p(\mathbf{\Phi})  \mathbf{H}_{1},
\end{aligned}
\end{equation} and $\widetilde{\mathbf{W}}_{p}\triangleq\mathbf{W}_{p}\mathbf{X}^{H}\in \mathbb{C}^{K \times M}$ is the noise matrix after pilots' removal. Each $(k,m)$-th entry of $\mathbf{Z}_{p}$ with $k=1,2,\ldots,K$ and $m=1,2,\ldots,M$ is obtained as
\begin{equation}\label{scalar_decomp}
[\mathbf{Z}_{p}]_{k,m}=\sum_{n=1}^{N} [\mathbf{H}_{2}]_{k,n} [\mathbf{H}_{1}]_{n,m} [\mathbf{\Phi}]_{p,n},
\end{equation}
where $[\mathbf{H}_{1}]_{n,m}$, $[\mathbf{H}_{2}]_{k,n}$, and $[\mathbf{\Phi}]_{p,n}$ denote the $(n,m)$-th entry of $\mathbf{H}_{1}$, $(k,n)$-th entry of $\mathbf{H}_{2}$, and the $(p,n)$-th entry of $\mathbf{\Phi}$, respectively, with $n=1,2,\ldots,N$.

Resorting to the PARAFAC decomposition \cite{harshman1994parafac,bro2003new,ten2002uniqueness,roemer2008closed}, each matrix $\mathbf{Z}_{p}$ in \eqref{scalar_decomp} out of the $P$ in total can be represented using three matrix forms. These matrices form the horizontal, lateral, and frontal slices of the tensor composed of \eqref{scalar_decomp}. The unfolded forms of the mode-1, mode-2, and mode-3 of $\mathbf{Z}_{p}$'s are expressed as follows:
\begin{equation}\label{mode_1}
\text{Mode-1:} \quad {\mathbf{Z}_\alpha=(\mathbf{H}_1^T \circ   \mathbf{\Phi})\mathbf{H}_2^T}\in\mathbb{C}^{PM \times K},
\end{equation}
\begin{equation}\label{mode_2}
\text{Mode-2:}  \quad {\mathbf{Z}_\beta=( \mathbf{\Phi} \circ \mathbf{H}_2)\mathbf{H}_1}\in\mathbb{C}^{KP \times M},
\end{equation}
\begin{equation}\label{mode_3}
\quad \text{Mode-3:} \quad {\mathbf{Z}_\gamma=( \mathbf{H}_2 \circ  \mathbf{H}_1^T)  \mathbf{\Phi}^T}\in\mathbb{C}^{MK \times P}.
\end{equation}
Clearly, each of the postprocessed noisyless received channel matrices $\mathbf{Z}_\alpha$, $\mathbf{Z}_\beta$, and $\mathbf{Z}_\gamma$ is expressed as a matrix product of a Khatri-Rao matrix product factor and a single  matrix, which it will be shown in the sequel that facilitates the channel estimation problem. In the following section, we present an ALS-based algorithm \cite{kroonenberg1980principal} for iteratively estimating the channel matrices $\mathbf{H}_{1}$ and $\mathbf{H}_{2}$.

\section{Proposed Iterative Channel Estimation}\label{sec:channel_est}
In this section, we present an iterative algorithm for the estimation of all involved channel matrices at the reception side, and discuss its feasibility conditions. 

\subsection{Proposed Algorithm}
We start with the definition of the following three-dimensional matrix $\widetilde{\mathbf{Z}}\in \mathbb{C}^{K \times M \times P}$:
\begin{equation}\label{noise_term}
\widetilde{\mathbf{Z}}\triangleq\mathbf{Z}+\widetilde{\mathbf{W}},
\end{equation}
which includes all $P$ matrices $\widetilde{\mathbf{Z}}_p$ in \eqref{widetilde_Z_p} in its third dimension. Similarly, tensor $\widetilde{\mathbf{W}} \in \mathbb{C}^{K \times M \times P}$ is the AWGN that incorporates all $P$ matrices $\widetilde{\mathbf{W}}_p$. In \eqref{noise_term}, tensor $\mathbf{Z}$ represents the noiseless version of $\widetilde{\mathbf{Z}}$, which can be obtained using the unfolded forms \eqref{mode_1}, \eqref{mode_2}, and \eqref{mode_3}. We next present the algorithmic steps of the proposed ALS-based iterative channel estimation algorithm, which is summarized in Algorithm~1. 

%% Initilization step
\subsubsection{\textbf{First Step (Initialization)}}
To facilitate the considered iterative channel estimation method, we initiate the algorithm with the eigenvector matrices presented \cite{kroonenberg1980principal}. Particularly, $\widehat{\mathbf{H}}_1^{(0)}$ represents the eigenvector matrix corresponding to the $N$ non-zero eigenvalues of $\widetilde{\mathbf{Z}}_\beta^H \widetilde{\mathbf{Z}}_\beta$, where $\widetilde{\mathbf{Z}}_\beta\in \mathbb{C}^{KP \times M}$ is the mode-2 form of \eqref{noise_term}; this is \eqref{mode_2}'s noisy version. Similarly, $\widehat{\mathbf{H}}_2^{(0)}$ is the eigenvector matrix corresponding to the $N$ non-zero eigenvalues of $\widetilde{\mathbf{Z}}_\alpha^H \widetilde{\mathbf{Z}}_\alpha$ with $\widetilde{\mathbf{Z}}_\alpha\in \mathbb{C}^{PM \times M}$ being the mode-1 form of \eqref{noise_term}, which is actually the noisy version of \eqref{mode_1}.

%% Iteration and update
\subsubsection{\textbf{Second and Third Steps (Iterative Update)}}
The channels $\mathbf{H}_1$ and $\mathbf{H}_2$ are obtained in an iterative manner by alternatively minimizing conditional Least Squares (LS) criteria using tensor $\widetilde{\mathbf{Z}}$ for the postprocessed received signal given by \eqref{noise_term} and \eqref{mode_1}--\eqref{mode_3} for all $P$ matrices $\mathbf{Z}_p$. Starting with $\mathbf{H}_2$, we make use of the mode-1 unfolded form given by \eqref{mode_1} according to which ${\mathbf{Z}_\alpha=\mathbf{A}_1 \mathbf{H}_2^T}$ with $\mathbf{A}_1=\mathbf{H}_1^T\circ \mathbf{\Phi} \in \mathbb{C}^{PM \times N}$. 
%In order to estimate the channel $\mathbf{H}^{r}$, we use the mode-1 unfolded form given by \eqref{mode1_sim},
%\begin{equation}\label{mode1_sim}
%{\mathbf{Z}_\alpha=\mathbf{A}_1 \mathbf{H}_2^T}
%\end{equation}
%where $\mathbf{A}_1=\mathbf{H}_1^T\circ \mathbf{\Phi} \in \mathbb{C}^{PM \times N}$.
At the $i$-th algorithmic iteration, the $i$-th estimation for $\mathbf{H}_2$, denoted by $\widehat{\mathbf{H}}_2^{(i)}$, is obtained from the minimization of the following LS objective function: 
\begin{equation}\label{ls_hr}
\begin{aligned} J\left(\widehat{\mathbf{H}}_2^{(i)}\right)=\left\|\widetilde{\mathbf{Z}}'-\widehat{\mathbf{A}}_{1}^{(i-1)}\left(\widehat{\mathbf{H}}_2^{(i)}\right)^T\right\|_{F}^{2},
\end{aligned}
\end{equation}
where $\widetilde{\mathbf{Z}}'\in\mathbb{C}^{PM \times K}$ is a matrix-stacked form of \eqref{noise_term}'s tensor $\widetilde{\mathbf{Z}}$, and $\widehat{\mathbf{A}}_{1}^{(i-1)}=\widehat{\mathbf{H}}_2^{(i-1)}\circ\mathbf{\Phi}$. The solution of \eqref{ls_hr} is given by
\begin{equation}\label{ls_hr_solution}
\left(\widehat{\mathbf{H}}_2^{(i)}\right)^T=\left(\widehat{\mathbf{A}}_{1}^{(i-1)}\right)^{\dagger}\widetilde{\mathbf{Z}}'.
\end{equation}

In a similar way, for the $\mathbf{H}_1$ estimation, we use the mode-2 unfolded form \eqref{mode_2}, where ${\mathbf{Z}_\beta=\mathbf{A}_2\mathbf{H}_1}$ with $\mathbf{A}_2= \mathbf{\Phi} \circ\mathbf{H}_2 \in \mathbb{C}^{KP \times N}$, and formulate the following LS objective function for the $i$-th estimation for $\mathbf{H}_1$ denoted by $\widehat{\mathbf{H}}_1^{(i)}$:
\begin{equation}\label{ls_hs}
\begin{aligned} J\left(\widehat{\mathbf{H}}_1^{(i)}\right) &=\left\|\widetilde{\mathbf{Z}}''-\widehat{\mathbf{A}}_{2}^{(i)}\mathbf{H}_1 \right\|_{F}^{2},
\end{aligned}
\end{equation}
where $\widetilde{\mathbf{Z}}''\in\mathbb{C}^{KP \times M}$ is another matrix-stacked form of \eqref{noise_term}'s tensor $\widetilde{\mathbf{Z}}$, and $\widehat{\mathbf{A}}_{2}^{(i)}= \mathbf{\Phi}\circ\widehat{\mathbf{H}}_2^{(i)} $. The solution of \eqref{ls_hs} is easily obtained as
\begin{equation}\label{ls_hs_solution}
\begin{aligned}
\widehat{\mathbf{H}}_1^{(i)}=\left(\widehat{\mathbf{A}}_{2}^{(i)}\right)^{\dagger}\widetilde{\mathbf{Z}}''.
\end{aligned}
\end{equation}

\subsubsection{\textbf{Fourth Step (Iteration Stop Criterion)}}
The proposed iterative algorithm terminates when either the maximum number $I_{\rm max}$ of algorithmic iterations is reached or when between any two algorithmic iterations $i-1$ and $i$ holds $\|\widehat{\mathbf{H}}_{1}^{(i)}-\widehat{\mathbf{H}}_{1}^{(i-1)} \|_{F}^{2}\|\widehat{\mathbf{H}}_{1}^{(i)}\|_{F}^{-2}\leq\epsilon$ or $\|\widehat{\mathbf{H}}_{2}^{(i)}-\widehat{\mathbf{H}}_{2}^{(i-1)} \|_{F}^{2}\|\widehat{\mathbf{H}}_{2}^{(i)}\|_{F}^{-2}\leq\epsilon$ for $\epsilon$ being a very small positive real number.

\begin{algorithm}[htb]
	\textbf{Algorithm 1: Iterative Channel Estimation}
	\begin{enumerate}
		\item Initialize with a random feasible phase matrix $\Phi$ and $\widehat{\mathbf{H}}_1^{(0)}$ obtained from the $N$ non-zero eigenvalues of $\mathbf{Z}_\beta^H \mathbf{Z}_\beta$, and set algorithmic iteration $i=1$. 
		\item Obtain $\widehat{\mathbf{H}}_2^{(i)}$ using $\widehat{\mathbf{H}}_2^{(i)}=\left(\left(\widehat{\mathbf{A}}_{1}^{(i-1)}\right)^{\dagger}\widetilde{\mathbf{Z}}'\right)^T$.
		\item Obtain $\widehat{\mathbf{H}}_1^{(i)}$ using $\widehat{\mathbf{H}}_1^{(i)}=\left(\widehat{\mathbf{A}}_{2}^{(i)}\right)^{\dagger}\widetilde{\mathbf{Z}}''$.
		\item If $\frac{\|\widehat{\mathbf{H}}_{1}^{(i)}-\widehat{\mathbf{H}}_{1}^{(i-1)} \|_{F}^{2}}
		{ \|\widehat{\mathbf{H}}_{1}^{(i)} \|_{F}^{2}} \leq \epsilon$ or $\frac{\|\widehat{\mathbf{H}}_{2}^{(i)}-\widehat{\mathbf{H}}_{2}^{(i-1)} \|_{F}^{2}}
		{ \|\widehat{\mathbf{H}}_{2}^{(i)} \|_{F}^{2}} \leq \epsilon$, or $i=I_{\rm max}$, then end. Otherwise, set $i=i+1$ and go to step 2$)$.
	\end{enumerate}
\end{algorithm}

\subsection{Feasibility Conditions}
The iterative estimation of $\mathbf{H}_1$ and $\mathbf{H}_2$ using the ALS approach included in Algorithm~1 encounters a scaling ambiguity from the convergence point, which can be resolved with adequate normalization \cite{rong2012channel,lioliou2009performance}. In addition, to ensure that Algorithm~1 yields a solution, some of the system parameters need to meet necessary and sufficient conditions for system identifiability \cite{rong2012channel}. For the considered RIS-assisted multi-user MISO communication system, the feasibility conditions include the following inequalities for the system parameters:
\begin{equation}\label{eq:feasibility}
M \geq N \,\,\, \text{and} \,\,\, \operatorname{min}\left(N, K\right) \geq N.
\end{equation}
Hence, for the PARAFAC model in \eqref{scalar_decomp} to hold, it should be $M,K \geq N$ and the number of training phase configurations $P$ should be less or equal to $N$. For those cases, the triple $(\mathbf{\Phi}, \mathbf{H}_1, \mathbf{H}_2)$ is unique up to some permutation and scaling ambiguities. The detailed feasibility proof can be derived using \cite{rong2012channel}, which is however omitted here due to space limitation.

\begin{remark}
It is expected, in practice, that RIS will contain large numbers $N$ of unit elements, and particularly, larger than the number $M$ of BS antenna elements and/or the number $K$ of single-antenna mobile users. In such cases, the feasibility conditions in \eqref{eq:feasibility} are not met, and Algorithm~1 needs to be deployed in the following way. The $N$-element RIS should be first partitioned in groups of non-overlapping cells, such that each group's number of unit elements meets the feasibility conditions (i.e., less or equal to $M$ and $K$). Then, Algorithm~1 can be used for each group to estimate portions of the intended channels that can be finally combined to derive the desired channel estimation. This general case will be explicitly treated in the extended version of this work.      
%Although the number of $N$-elements in the each RIS cell is limited by $M$ and $K$, resulting by the PARAFAC method, multiple RIS cells can construct a large RIS with the number of all elements that is far larger than $M$ and $K$. In other words, a large RIS can be grouped into many independent cells, and each cells consists of $N$ elements that can be independently estimated by leveraging the proposed algorithm.
\end{remark}

%%% Simulation part
\section{Performance Evaluation Results}\label{sec:simulation}
In this section, we present computer simulation results for the performance evaluation of the proposed channel estimation algorithm. We have particularly simulated the Normalized Mean Squared Error (NMSE) of our channel estimator using the metrics $\|\mathbf{H}_{1}-\widehat{\mathbf{H}}_{1}\|^{2}\|\mathbf{H}_{1}\|^{-2}$ and $\|\mathbf{H}_{2}-\widehat{\mathbf{H}}_{2}\|^{2}\|\mathbf{H}_{2}\|^{-2}$. The scaling ambiguity of Algorithm~1 has been removed by normalizing the first column of $\mathbf{H}_1$, such that it has all one elements. For $\mathbf{\Phi}$, we have considered the discrete Fourier transform matri, which satisfies $\mathbf{\Phi}^H\mathbf{\Phi}=\mathbf{I}_{N}$ and is suggested as a good choice for ALS in \cite{rong2012channel}. All NMSE curves were obtained after averaging over $2000$ independent Monte Carlo channel realizations. For Algorithm~1, we have used $\epsilon=10^{-5}$ and $I_\text{max}=20$ in all presented NMSE performance curves.

\begin{figure} \vspace{-1mm}
	\begin{center}
		\includegraphics[width=0.44\textwidth]{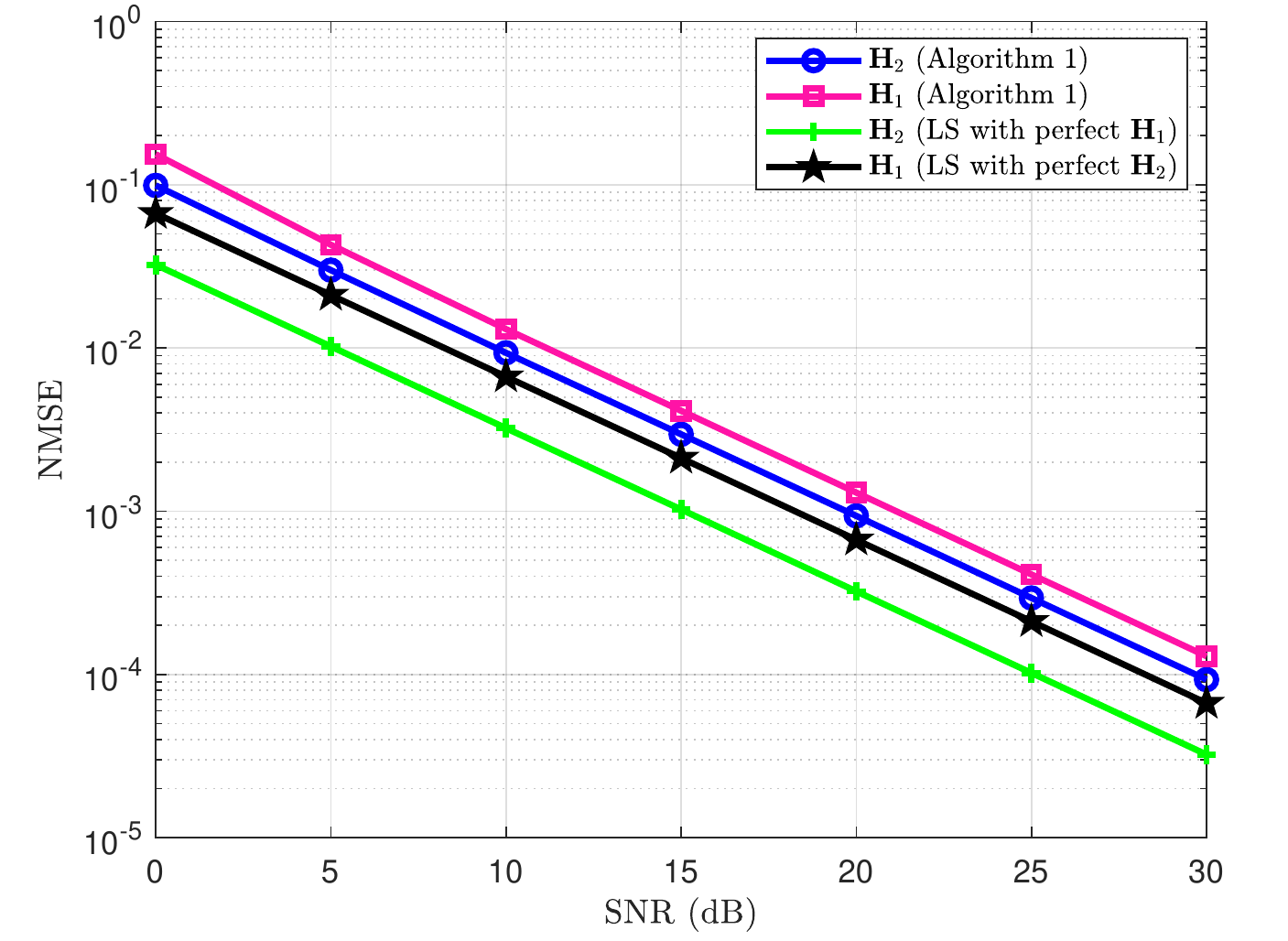}  \vspace{-2mm}
		\caption{NMSE performance comparisons between the proposed algorithm and conventional LS estimation versus the SNR in dB for $M=32$, $K=16$, $T=32$, $N=16$, and $P=16$.}
		\label{fig:com_ls2} \vspace{-6mm}
	\end{center}
\end{figure}

\begin{figure}\vspace{-1mm}
	\begin{center}
		\centerline{\includegraphics[width=0.44\textwidth]{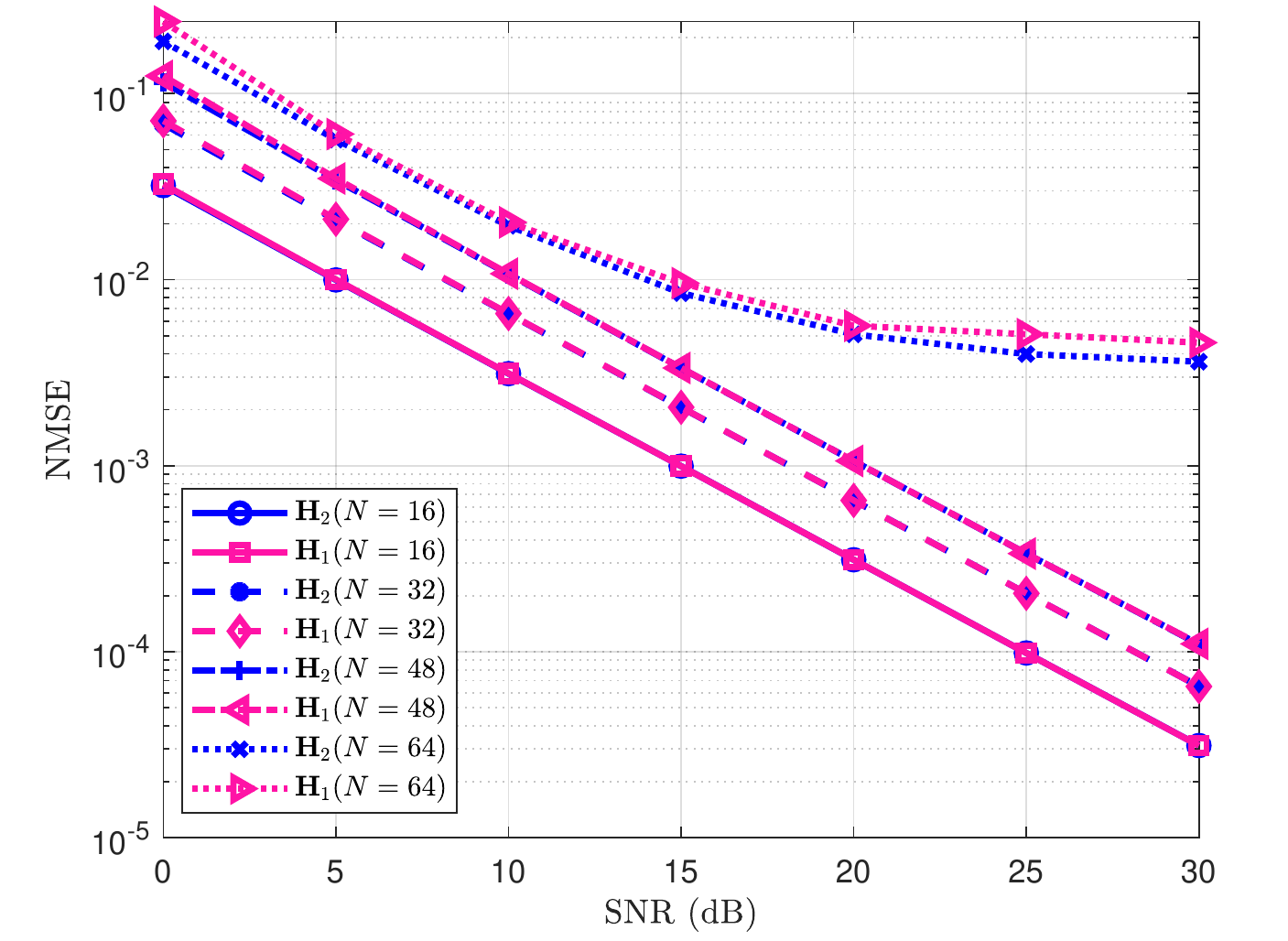} } \vspace{-2mm}
		\caption{NMSE performance of the proposed algorithm versus the SNR in dB for $M=K=T=64$, $P=16$, and various values of $N$.}
		\label{fig:n_hr_hs} \vspace{-6mm}
	\end{center}
\end{figure}

The performance comparisons between the proposed channel estimation Algorithm~1 and conventional LS estimation is illustrated in Fig$.$~\ref{fig:com_ls2} as a function of the Signal plus Noise Ratio (SNR) for $M=32$, $K=16$, $T=32$, $N=16$, and $P=16$. In this figure, the green NMSE curve denotes the LS estimation for $\mathbf{H}_2$ given that $\mathbf{H}_1$ is perfectly known, while the black curve refers to the LS estimation for $\mathbf{H}_1$ when $\mathbf{H}_2$ is perfectly available. The pink and blue NMSE curves are the estimations for $\mathbf{H}_1$ and $\mathbf{H}_2$ provided by Algorithm~1 that requires no a priori information. As shown, the proposed iterative algorithm achieves comparable performance with the benchmark scheme. Specifically, there is around $4$dB gap in the estimation of $\mathbf{H}_1$ between the proposed algorithm and LS estimation with the a priori $\mathbf{H}_1$ knowledge, while the gap for the $\mathbf{H}_2$ estimation between the two methods is reduced to about $2.5$dB. This difference happens due to the fact that $\mathbf{H}_2$ is a square matrix with less unknown elements compared with $\mathbf{H}_1$.
In Fig$.$~\ref{fig:n_hr_hs}, we have used the parameters' setting $M=K=T=64$, $P=16$, and various values of $N$ for illustrating the NMSE performance of Algorithm~1 as a function of the SNR. It is evident that there exists an increasing performance loss when increasing the number $N$ of RIS unit elements. A large $N$ results in a large number of rows and columns of $\mathbf{H}_1$ and $\mathbf{H}_2$, respectively, which requires larger computational complexity for ALS-based estimation.

\begin{figure} \vspace{-1mm}
	\begin{center}
		\includegraphics[width=0.44\textwidth]{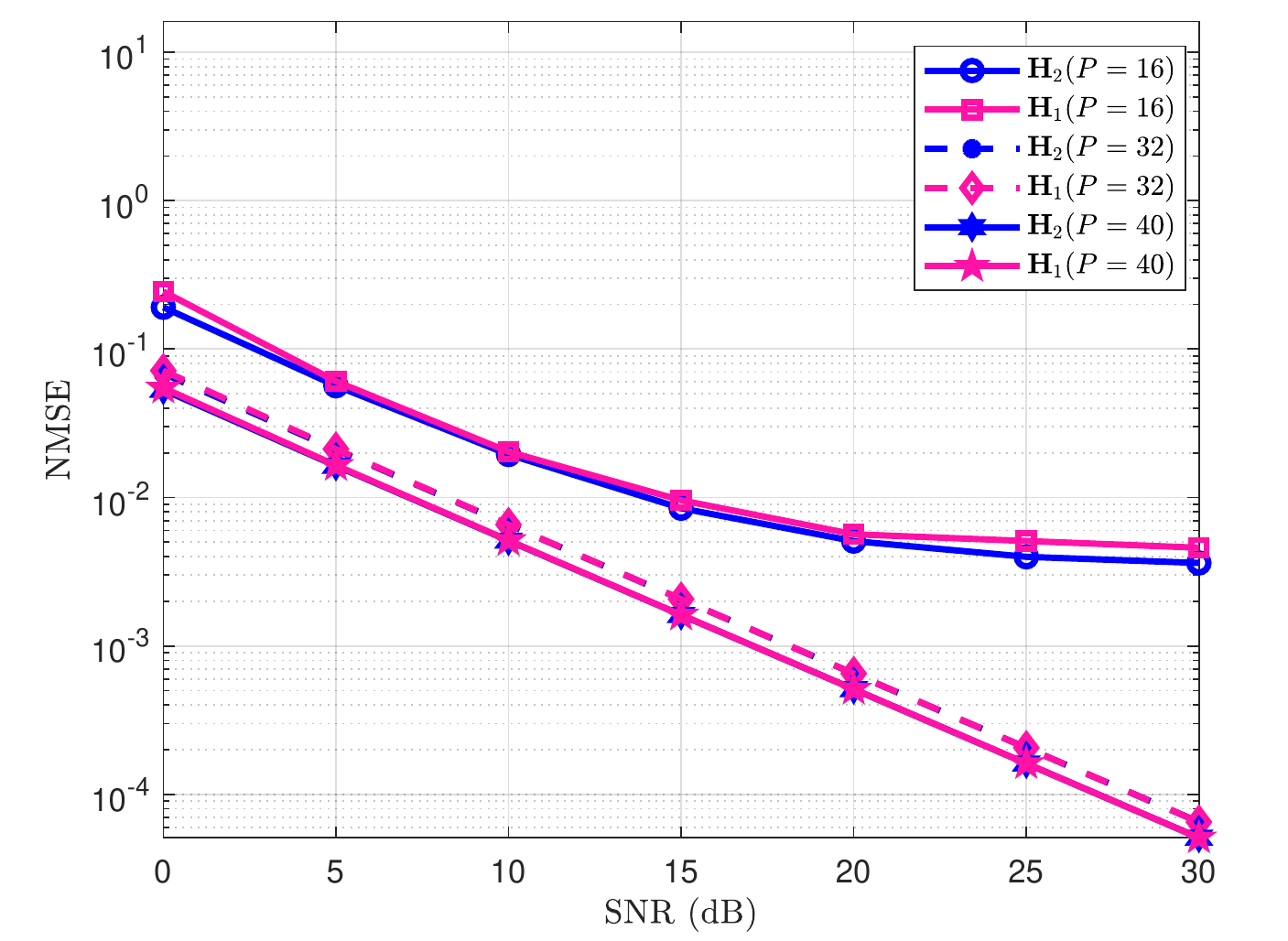}  \vspace{-2mm}
		\caption{NMSE performance of the proposed algorithm versus the SNR in dB for $M=K=T=N=64$ and various values of $P$.}
		\label{fig:p_hr_hs} \vspace{-6mm}
	\end{center}
\end{figure}

The impact of the number $P$ for the RIS training phase configurations in the NMSE performance of the proposed algorithm is investigated in Fig$.$~\ref{fig:p_hr_hs} as a function of the SNR for $M=K=T=N=64$. It is shown that increasing $P$ improves NMSE with $P=40$ yielding the best performance. This happens because with increasing $P$, the training set increases benefiting channel estimation. However, the performance improvement between adjacent NMSE curves decreases as $P$ increases. It is noted that larger $P$ results in higher computational complexity, and that for the considered parameter setting it must hold $P\leq64$ according to the feasibility conditions of Algorithm~1. 

\section{Conclusion}\label{sec:conclusion}
In this paper, we proposed an iterative ALS-based channel estimation method for RIS-assisted multi-user MISO communication systems that capitalizes on the PARAFAC decomposition of the received signal model. At each algorithmic step, the proposed algorithm obtains the estimations for the channel between BS and RIS, as well as the channels between RIS and users in closed form. We also investigated the feasibility conditions for the proposed iterative algorithm. Our simulation results showed that the designed channel estimation outperforms a benchmark LS scheme, and that the numbers of RIS unit elements and training symbols exert a significant effect on the proposed algorithm.

\newpage

\bibliographystyle{IEEEtran}
\bibliography{strings}
\vspace{12pt}

\end{document}